\documentclass[aps, prb, preprint, superscriptaddress, a4paper, 11pt, floatfix,usenames,dvipsnames]{revtex4-1}

\usepackage{graphicx}
\usepackage{epstopdf}
\usepackage[multi-part-units = single]{siunitx}
\usepackage{amsmath}
\usepackage{lineno}
\usepackage[export]{adjustbox}
\usepackage{comment}
\usepackage{xcolor}
\usepackage[overload]{textcase} 

\usepackage{silence}
\usepackage{colortbl}

\newcommand{\TV}[1]{}

\newcommand{\Chen}[1]{\textcolor{blue}{#1}}

\newcommand{\magenta}[1]{\textcolor{magenta}{#1}}

\newcommand{\Oren}[1]{\textcolor{orange}{#1}}
\newcommand{\Matan}[1]{\textcolor{OliveGreen}{#1}}

\newcommand{\gt}{$g^{(2)}$\,}
\newcommand{\avg}[1]{\langle {#1} \rangle}

\newcommand{\Cov}[1]{\text{Cov}({#1})}
\newcommand{\gtm}{g^{(2)}}
\newcommand{\beq}{\begin{equation}}
\newcommand{\eeq}{\end{equation}}

\usepackage{float}

\usepackage[bookmarks=True,bookmarksopen=True]{hyperref} 
\hypersetup {
	unicode=false,          
	pdftoolbar=true,        
	pdfmenubar=true,        
	pdffitwindow=false,     
	pdfstartview={FitH},    
	colorlinks=true,       
	linkcolor=BlueViolet,          
	citecolor=BlueViolet,        
	filecolor=Black,      
	urlcolor=Black           
}

\graphicspath{{./main_figures/}}  

\usepackage{geometry}
\usepackage{soul}
\pdfpageattr {/Group << /S /Transparency /I true /CS /DeviceRGB>>}

\begin{document}

\newgeometry{left=3cm,right=3cm,top=2.5cm,bottom=2.5cm}
\title{Measuring and controlling the birth of quantum attosecond pulses}

\author{Matan Even Tzur\textsuperscript{\textdagger}}
\affiliation{Department of Physics, Technion—Israel Institute of Technology, Haifa, Israel}
\affiliation{Solid State Institute, Technion—Israel Institute of Technology, Haifa, Israel}
\affiliation{Helen Diller Quantum Center, Technion—Israel Institute of Technology, Haifa, Israel}

\author{Chen  Mor\textsuperscript{\textdagger}}
\affiliation{Department of Physics of Complex Systems, Weizmann Institute of Science, Rehovot, Israel}

\author{Noa Yaffe}
\affiliation{Department of Physics of Complex Systems, Weizmann Institute of Science, Rehovot, Israel}

\author{Michael Birk}
\affiliation{Solid State Institute, Technion—Israel Institute of Technology, Haifa, Israel}
\affiliation{Helen Diller Quantum Center, Technion—Israel Institute of Technology, Haifa, Israel} 
\affiliation{ The Russell Berrie Nanotechnology Institute , Technion—Israel Institute of Technology, Haifa, Israel}

\author{Andrei Rasputnyi}
\affiliation{Max Planck Institute for the Science of Light, Erlangen, Germany
}
\affiliation{Friedrich–Alexander Universität Erlangen–Nürnberg, Erlangen, Germany}

\author{Omer Kneller}
\affiliation{Department of Physics of Complex Systems, Weizmann Institute of Science, Rehovot, Israel}

\author{Ido Nisim}
\affiliation{Department of Physics, Technion—Israel Institute of Technology, Haifa, Israel}
\affiliation{Solid State Institute, Technion—Israel Institute of Technology, Haifa, Israel}
\affiliation{Helen Diller Quantum Center, Technion—Israel Institute of Technology, Haifa, Israel}

\author{Ido Kaminer}
\affiliation{Solid State Institute, Technion—Israel Institute of Technology, Haifa, Israel}
\affiliation{Helen Diller Quantum Center, Technion—Israel Institute of Technology, Haifa, Israel} 
\affiliation{Department of Electrical and Computer Engineering, Technion—Israel Institute of Technology, Haifa, Israel}

\author{Michael Krüger}
\affiliation{Department of Physics, Technion—Israel Institute of Technology, Haifa, Israel}
\affiliation{Solid State Institute, Technion—Israel Institute of Technology, Haifa, Israel}
\affiliation{Helen Diller Quantum Center, Technion—Israel Institute of Technology, Haifa, Israel}

\author{$ \big( \mid$ Nirit Dudovich, Oren Cohen $\rangle$ + $\mid$ Oren Cohen, Nirit Dudovich $\rangle \big)$}
\affiliation{Department of Physics, Technion—Israel Institute of Technology, Haifa, Israel}
\affiliation{Solid State Institute, Technion—Israel Institute of Technology, Haifa, Israel}
\affiliation{Helen Diller Quantum Center, Technion—Israel Institute of Technology, Haifa, Israel} 
\affiliation{Department of Physics of Complex Systems, Weizmann Institute of Science, Rehovot, Israel}
\affiliation{Guangdong Technion-Israel Institute of Technology, Shantou, Guangdong 515063, China   \\ \textsuperscript{\textdagger} These authors contributed equally to this work}

\begin{abstract}
The generation and control of extreme ultraviolet (XUV) radiation by high harmonic generation (HHG) have advanced ultrafast science, providing direct insights into electron dynamics on their natural time scale. Attosecond science has established the capability to resolve ultrafast quantum phenomena in matter by characterizing and controlling the classical properties of the high harmonics. Recent theoretical proposals have introduced novel schemes for generating and manipulating XUV HHG with distinct quantum features, paving the way to attosecond quantum optics. In this work, we transfer fundamental concepts in quantum optics into attosecond science. 
By driving the HHG process with a combination of an infrared bright squeezed vacuum (BSV, a non-classical state of light), and a strong coherent field, we imprint the quantum correlations of the input BSV onto both the ultrafast electron wavefunction and the harmonics' field.  Performing in-situ HHG interferometry provides an insight into the underlying sub-cycle dynamics, revealing squeezing in the statistical properties of one of the most fundamental strong-field phenomena -- field induced tunneling. Our measurement allows the reconstruction of the quantum state of the harmonics through homodyne-like tomography, resolving correlated fluctuations in the harmonic field that mirror those of the input BSV. By controlling the delay between the two driving fields, we manipulate the photon statistics of the emitted attosecond pulses with sub-cycle accuracy. The ability to measure and control quantum correlations in both electrons and XUV attosecond pulses establishes a foundation for attosecond electrodynamics, manipulating the quantum state of electrons and photons with sub-cycle precision. 
\end{abstract}

\maketitle

\section{\NoCaseChange{Introduction}}

Attosecond science has revolutionized our ability to capture and control fundamental ultrafast phenomena, revealing electron dynamics on their natural time scale \cite{Villeneuve2018}. The foundation and driving force of this field is the generation of XUV attosecond pulses via a strong-field light-matter interaction, known as high harmonic generation (HHG) \cite{Ferray1988,Li1989}. Over the past two decades XUV HHG spectroscopy has resolved, with unprecedented resolution, a large range of fundamental quantum phenomena -- from field induced tunneling \cite{Pedatzur2015} \cite{he2010interference} or multi-electron dynamics in atoms \cite{shiner2011probing}
,  to charge migration in molecules \cite{ChargeMigration} and topological dynamics in solids \cite{TopologicalHHG}. 
These experiments share one common characteristic -- the \textit{classical} 
properties of attosecond pulses served as a probe to resolve or manipulate the \textit{quantum} dynamics of electronic wavefunctions. While the measurement and control of the classical properties of attosecond pulses and their associated XUV high harmonics are established,\cite{Gariepy2014,Fleischer2014, huang2018polarization} their quantum characteristics remain neither measured nor controlled.

Quantum light \cite{loudon2000quantum}, predominantly produced in the visible to infrared spectral range, plays a critical role in quantum information processing \cite{Braunstein2005}, precision metrology \cite{Aasi2013}, and photonic quantum technologies \cite{PhotonicQuantumTech}. While most quantum light sources are relatively weak, the generation of bright squeezed vacuum (BSV) can extend quantum optics into the highly nonlinear regime \cite{Chekhova2015}. BSV is generated by high-gain spontaneous parametric amplification of vacuum fluctuations, producing entangled photon pairs while ‘squeezing’ quantum noise in one of the field's quadratures. It is a superposition of even-photon-number states with a zero mean field (⟨E(t)⟩ = 0) and correlated (squeezed) electric field fluctuations\cite{Sharapova2020}. 
Recent experiments employing BSV to drive HHG \cite{Rasputnyi2024, Lemieux2024} and multiphoton photoemission \cite{Heimerl2024} reveal that the photon number statistics of UV photons and photoelectrons are reminiscent of those in the non-classical BSV.
Other experiments explored the quantum optical signatures of the HHG mechanism when driven by coherent state light, studying its influence on the outgoing infrared driving field \cite{Lewenstein2021} and visible harmonics \cite{Theidel2024}. 
These pioneering experiments resolved quantum photon and electron statistics, revealing the role of quantum optics in strong-field light-matter interactions. However, attosecond-scale dynamics, encoded within the quantum states of the emitted photons and electrons, have remained unexplored.

Recent theoretical studies suggest that applying quantum optics concepts to attosecond pulses and their corresponding XUV high harmonics opens the door to a new class of quantum phenomena where the quantum properties of light manipulate matter, or are controlled by it, 
on a sub-cycle time scale \cite{Gorlach2020,Lewenstein2021, Gorlach2023, Harrison2023, EvenTzur2023, Tzur2024, Sloan2023, Lange2024, Yi2024, RiveraDean2024}. It has been predicted that HHG can generate entangled XUV photon pairs \cite{Sloan2023}, massively entangled \cite{Lange2024,Yi2024}, and squeezed light states\cite{Tzur2024,RiveraDean2024}. Moreover, it has been proposed that the quantum states of high harmonics could unveil many-body electronic correlations in quantum systems, such as correlated materials \cite{Lange2024} and superradiating ensembles \cite{pizzi2023light}. These remarkable opportunities pose a critical challenge: can the quantum state of attosecond pulses and their associate XUV high harmonics be characterized and controlled? Furthermore, what is the link between the quantum state of these attosecond pulses and the underlying strong-field dynamics that lead to their generation?


Here, we demonstrate the generation of an attosecond pulse train in the XUV regime exhibiting controllable quantum properties, by manipulating the correlations of the electronic wavefunctions with sub-cycle precision. Combining a strong coherent field centered at $800$nm with a BSV centered at $1600$nm, we transfer the statistical properties of the quantum driving field into the sub-cycle dynamics of the field driven wavepacket. The wavepacket dynamics are subsequently imprinted onto the quantum properties of the emitted attosecond pulses. The two-color field acts as a sub-cycle interferometer \cite{Pedatzur2015}, revealing the attosecond scale correlations of the electronic wavefunction and providing a direct access to phase information. The interferometric nature of our scheme enables homodyne-like quantum state tomography, allowing the reconstruction of the Wigner function of the XUV light. Our study transfers quantum optics to new extremes by enabling unique phase-resolved correlation measurements on attosecond time scales, opening new opportunities for experimental realizations of attosecond quantum electrodynamics (QED). 

	\begin{figure}[!htb]
		\begin{center}
		
		\centering{\includegraphics*[width=1\columnwidth]{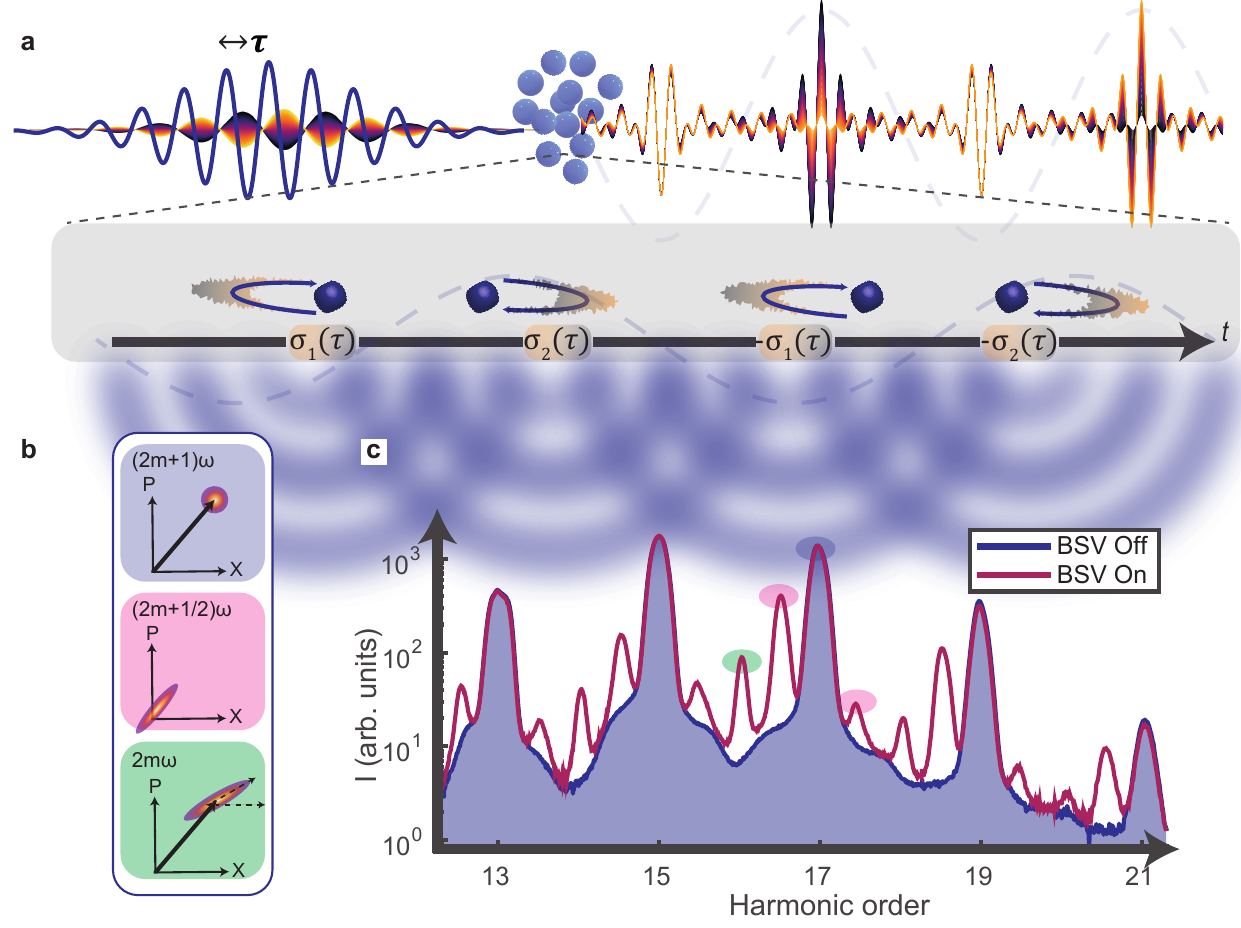}}
		\caption{ {\bf Generation of squeezed attosecond XUV pulses}. {\bf a},  HHG is driven by the combination of a strong coherent field of frequency $\omega$ and a weak BSV field of frequency $\omega$/2, emitting train of quantum attosecond pulses. The $\omega$-$\omega$/2 geometry realizes a temporal analogue of a 4-slit interferometer: emission of four fluctuating attosecond pulses per period of the $\omega/2$ field interfere to yield comb of integer and half-integer squeezed harmonics.  {\bf b}, Schematic illustrations of phase space diagrams associated with different harmonic orders. Odd harmonics approximately occupy a coherent state, half-integer harmonics exhibit squeezed fluctuations, and even harmonics exhibiting phase-space displacement and squeezed fluctuations. {\bf c}, Experimentally resolved HHG spectra generated by a coherent field only (blue) and the two-color field, composed of coherent and BSV sources (red). The perturbative field leads to the appearance of half-integer (pink) and even (green) harmonics.}
			\label{firstFigure}
		\end{center}
	\end{figure}

\section{\NoCaseChange{Attosecond quantum interferometry}}

Controlling and characterizing the quantum properties of attosecond pulses imposes a significant challenge. Attosecond pulses are generated during strong-field light-matter interaction, involving large photon numbers, while quantum light commonly involves low photon numbers. We overcome this challenge by combining a strong coherent beam centered at $800$nm with a weaker  BSV beam centered at $1600$nm. The coherent field governs the strong-field dynamics, while the BSV field acts as a perturbation, imprinting its non-classical characteristics onto the correlations and statistical properties of the generated attosecond pulses. 


HHG in gaseous media adheres to the well-established three-step model\cite{corkum2007attosecond}: an electron tunnels out under the influence of a strong driving laser field, accelerates freely in the continuum, and then recombines with the parent ion, emitting an XUV attosecond pulse. When driven solely by a fundamental $800$nm field ($\omega$), the symmetry between consecutive half cycles ensures a constructive interference at odd harmonics and destructive interference at even harmonics\cite{BenTal1993}. Introducing a $1600$nm perturbation with a frequency \(\omega/2\) breaks this symmetry, imposing a complex phase shift to the electronic semi-classical actions \(\sigma_j =\alpha_j +i \beta_j\) \cite{dahlstrom2011quantum}, associated with each fundamental half cycle \(j\). The real part of the perturbation, \(\alpha_j\), is dominated by a phase accumulated along the trajectory of the electron, while its imaginary part, \(\beta_j\), represents the perturbation of the field's amplitude, dominated by the tunneling mechanism. For a sufficiently weak $1600$nm field, these shifts are approximately linear with the perturbation amplitude. The perturbation exhibits a periodicity spanning four half cycles of the coherent field, with symmetry that dictates the relationships among the four complex  phases: 
\(\sigma_1, \sigma_2, \sigma_3 = -\sigma_1, \sigma_4 = -\sigma_2\) (Fig. \ref{firstFigure}a). This configuration serves as a four-ports in-situ interferometer, mapping the complex phase perturbation into the emergence of new frequency components -- half-integer harmonics ($2N\pm{\frac{1}{2}}$) and even harmonics ($2N$) -- while also modifying the odd harmonics ($2N+1$) -- as described by (SI section V): 
\begin{equation}
I_N \propto 
\begin{cases} 
    \left| \cos( \sigma_1) + \cos( \sigma_2) \right|^2, & \text{2N+1}, \\
    \left| \cos( \sigma_1) - \cos( \sigma_2) \right|^2, & \text{2N}, \\
    \left| i\sin( \sigma_1) + \sin( \sigma_2) \right|^2, & 
    \text{2N} + \frac{1}{2}, \\
    \left| i\sin( \sigma_1) - \sin( \sigma_2) \right|^2, & 
    \text{2N} - \frac{1}{2}.
\end{cases}
\label{eq:I}
\end{equation}


Notably, in contrast to previous works \cite{Pedatzur2015,luu2018observing, worner2010following,he2010interference, dahlstrom2011quantum}, in our experiment $\sigma_1$ and $\sigma_2$ as well as the harmonic intensities are inherently stochastic variables. The physical origin of their fluctuations is the non-classical noise of the BSV. In the absence of the BSV perturbation, the odd harmonics occupy coherent states \cite{rivera2022strong}, a property that remains approximately valid for a sufficiently weak BSV perturbation. In contrast, a perturbative analysis of Eq. 1 with respect to $\sigma_1$ and $\sigma_2$ shows that half-integer harmonics are approximately linear with the perturbative field and hence are anticipated to closely follow the BSV statistics, exhibiting vanishing displacement in phase space and squeezed fluctuations \cite{Tzur2024} (Fig. \ref{eq:I}b). Even harmonics are quadratic with the perturbation and are thus expected to occupy more complex states, exhibiting non-Gaussian fluctuations.
Equation \ref{eq:I} maps the quantum noise of the electronic wavefunction into measurable statistics of the harmonic intensities, and vice versa. As we shall show below, we utilize this mapping and its dependence on the delay between the coherent and BSV beams, to explore the quantum correlations of the electronic wavefunctions, and reconstruct the quantum states of the XUV high harmonics.

\section{\NoCaseChange{Photon statistics}}

Photon statistics characterize light by describing its photon number distribution and correlations, serving as one of the key methods for identifying the non-classical fingerprint of light. We first explore the photon statistics of the HHG radiation, dictated by the interplay between the strong-field interaction and the BSV perturbation. While a previous study measured photon statistics and second-order coherence of UV harmonics from solids \cite{Lemieux2024}, here we resolve these properties of gas-phase HHG in the attosecond regime and XUV spectral range.  As expected for BSV, the perturbation of this source is characterized by super-Poissonian photon statistics, exhibiting a long tail extending towards high photon numbers (SI section II). We resolve the photon statistics of the high harmonics by conducting single-shot measurements of the XUV spectrum, obtaining the complete photon statistics at a fixed two-color delay (SI section III).
  
	\begin{figure}[!htb]
		\begin{center}
		
		\centering{\includegraphics*[width=0.95\columnwidth]{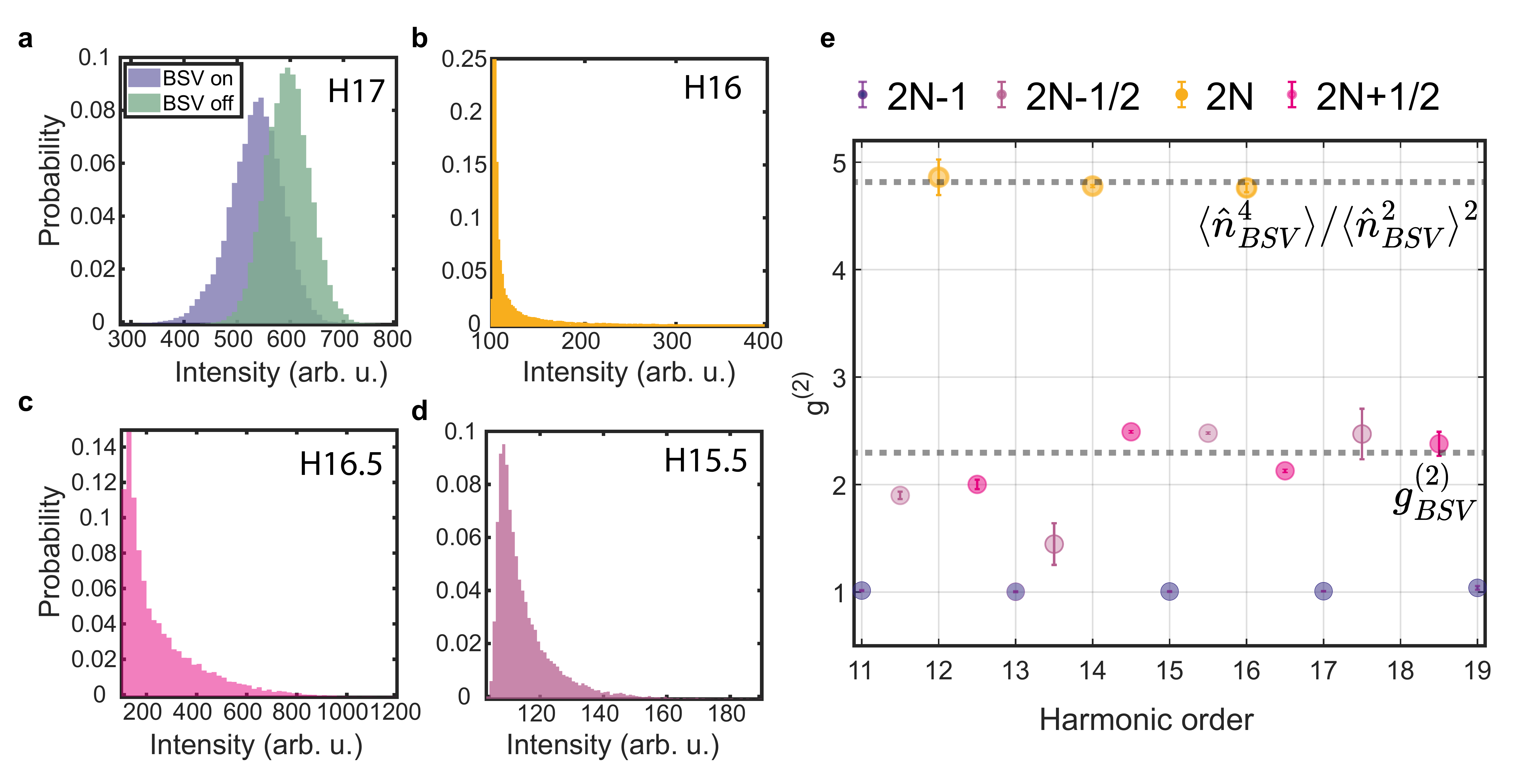}}
		\caption{{\bf Photon statistics of the XUV harmonics for a fixed two-color delay.} Intensity distributions of four types of harmonics: \textbf{a}, $2N+1$ (17), \textbf{b}, 2N (16), \textbf{c}, $2N + \frac{1}{2}$ (16.5), and \textbf{d}, $2N-\frac{1}{2}$ (15.5) . \textbf{e}, Second order coherence (\gt) as a function of harmonic order. Odd harmonics exhibit Poissonian statistics, with $\gtm\approx 1$. Half-integer and even harmonics exhibit a long-tailed distribution, with \gt values clustering around 2.3 and 4.8, respectively. These values correspond to \gt and the photon number kurtosis $\langle\hat{n}^4\rangle/\langle\hat{n}^2\rangle^2$ for the input BSV, as indicated by dashed gray lines in \textbf{e} (SI section II).} 
			\label{StaticFig}
		\end{center}
	\end{figure}

As described in Eq. \ref{eq:I}, the photon statistics of the harmonics group into four distinct families—odd, even, and two subcategories of half-integer harmonics (Figs \ref{StaticFig}a-d). Figure \ref{StaticFig}a presents the photon statistics of harmonic 17, which follows a Poissonian distribution in the absence of the BSV source, as expected when HHG is driven by a  classical coherent state field \cite{Gorlach2020,Lewenstein2021}. Introducing the BSV modifies this distribution, reducing the mean intensity and producing a broader, asymmetric profile, with the heavier tail extending towards lower photon numbers. This behavior reflects the anti-correlation between odd harmonic intensities and the perturbation amplitude, as predicted by Eq. \ref{eq:I}.
Figures \ref{StaticFig}b-d present the photon statistics of exemplary half-integer and even harmonics. These harmonic families exhibit photon statistics similar to those of the input BSV, with a long tail extending toward higher photon numbers.  
To quantify XUV intensity fluctuations, we calculate the second-order coherence, defined for a large photon number $\avg{\hat{n}}$ as $g^{(2)}= \langle\hat{n}^2-\hat{n}\rangle/\langle\hat{n}\rangle^2 \approx \langle\hat{n}^2\rangle/\langle\hat{n}\rangle^2 $,observing super-bunching ($\gtm > 2 $) in both half-integer and even harmonics (Fig. \ref{StaticFig}e). Half-integer harmonics cluster around $\gtm \approx2.3$, matching the input BSV value (SI section II), as they are mainly produced by a single-BSV-photon process \cite{MultiphotonMasha}. Even harmonics exhibit $g^{(2)}\approx 4.8$, matching the value of the photon number kurtosis $\langle\hat{n}^4\rangle/\langle\hat{n}^2\rangle^2$ for the input BSV, consistent with their generation by a two-BSV-photon process \cite{MultiphotonMasha}. 

\begin{figure}[!htb]
		\begin{center}
		
		\centering{\includegraphics*[width=1\columnwidth]{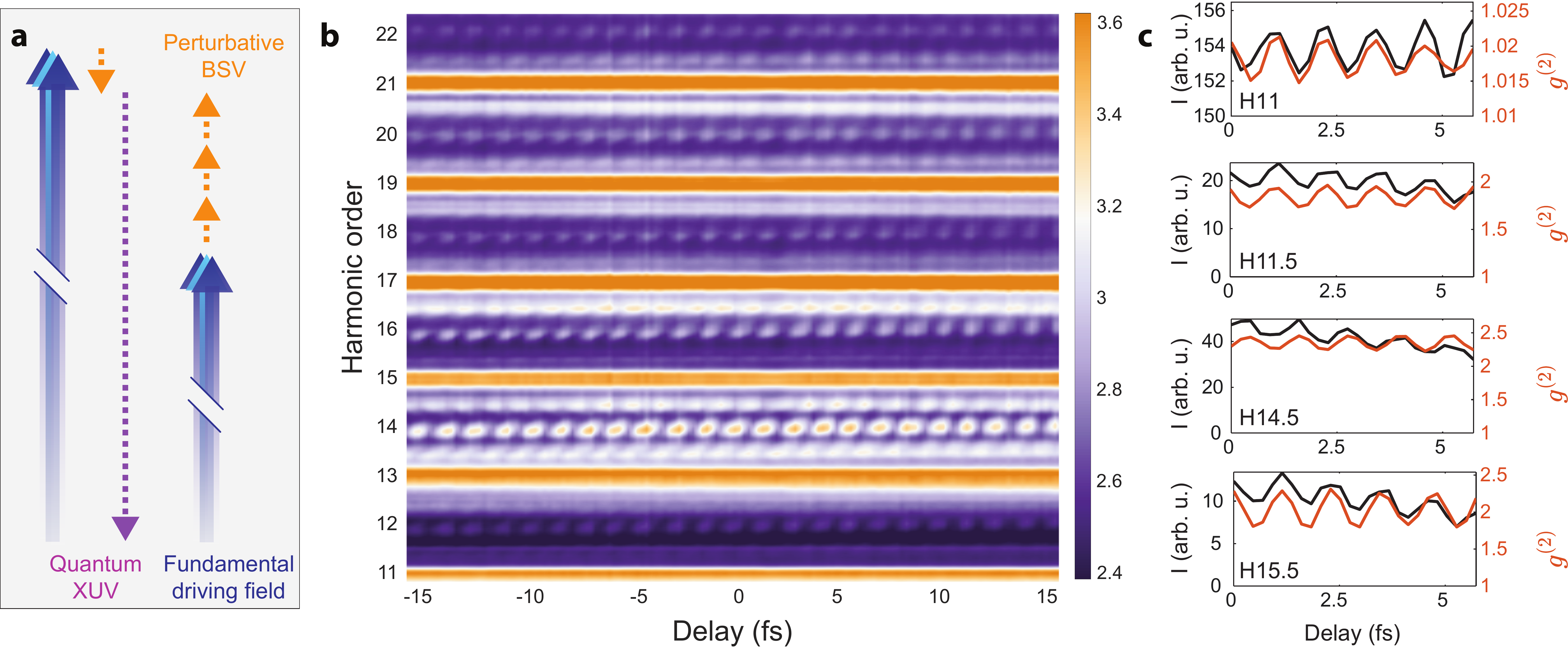}}
		\caption{{\bf Photon statistics of the XUV harmonics as a function of the two-color delay} {\bf a}, One and three BSV-photon (orange dashed arrows) processes interfere to generate half-integer harmonics. {\bf b}, Mean value of different optical frequencies as a function of the two-color delay. {\bf c}, \gt and mean oscillations of selected harmonics (11,11.5, 14.5 ,15.5)  as a function of the two-color delay (see SI section IV for the whole spectrum). For half-integer harmonics, the oscillations of \gt are in phase with the mean value oscillations. Notably, very weak, yet clearly resolved, \gt oscillations are observed in harmonic 11.
			}
			\label{DynamicFig}
		\end{center}
	\end{figure}

So far, we explored the photon statistics of the harmonics for a fixed two-color delay, revealing, for the first time, the mapping of squeezed vacuum photon statistics into the XUV harmonics. By measuring the single shot statistical distribution for each delay we track the sub-cycle dynamics that underlie this non-classical process. We observe periodic modulations of the mean intensity with a periodicity of $400$nm, corresponding to half the fundamental optical cycle (Fig. 3b). These oscillations originate from the modulations of sub-cycle trajectories, induced by the two-color field, periodically varying the phases of the temporal interferometer’s slits (Fig. \ref{firstFigure}). Statistical analysis shows that as we scan the two-color delay, \gt is modulated as well, in phase with the modulations of the mean value of the harmonics (Fig. \ref{DynamicFig}c). What is the origin of these oscillations? We address this question in time and then frequency domains, providing complementary insights about the sub-cycle quantum-optical nature of the HHG process. The BSV is uniquely characterized by a strongly fluctuating amplitude while maintaining a well defined phase. Such a perturbation leads to amplitude fluctuations in each of the four slits of the interferometer, which are mutually correlated and phase-locked with respect to the coherent driving field.  Consequently, their interference depends on the two-color delay, yielding the observed phase-locked mean intensity and \gt modulations. 

The perturbative nature of the BSV interaction allows us to describe the time domain modulations in the frequency domain, applying a photonic pathway picture. Expanding Eq.  \ref{eq:I} as a power series in the complex phases $\sigma_{1,2}$ describes the interaction as a sum over perturbative BSV photonic processes. Thus, we can view the mechanism as a frequency-domain interference of photonic pathways whose complex amplitudes depend on powers of $\sigma_{1,2}$. 
For example, half-integer harmonics oscillations originate from a combination of one-BSV-photon and three-BSV-photon processes  (Fig. \ref{DynamicFig}.a), as seen by expanding $\sin(\sigma_{1,2})\approx\sigma_{1,2}-\sigma_{1,2}^3/6$ in Eq. \ref{eq:I}. These pathways, $\sigma$ and $\sigma^3$, acquire different phases as we scan the two-color delay, leading to the modulations of the interference signal. Since the one and three-photon processes have distinct (but related) statistics,  \gt of their interference depend on their relative phase.
In addition, odd harmonic oscillations, described in Fig. \ref{DynamicFig}c, originate from the interference between a coherent state (zero-BSV-photons) and a four-BSV-photon process. Altogether, these results demonstrate how scanning the two-color delay uncovers interference among multiple photonic pathways, enabling precise control and characterization of the statistical properties of the emitted attosecond pulses.

\section{\NoCaseChange{Electron correlations and quantum state tomography}}


Next, we study the sub-cycle dynamics of the HHG electronic wavefunction, and their link to the quantum state of the emitted harmonics. According to Eq. \ref{eq:I}, the statistics of the recombining HHG electrons over four consecutive half-cycles—represented by the distributions of  $\sigma_{1,2}$—are mapped onto the joint intensity statistics of four adjacent harmonics (for instance, 14.5, 15, 15.5, and 16). Thus, by inverting Eq. \ref{eq:I}, we can extract two complex electron phases $\sigma_{1,2}\equiv\alpha_{1,2}+i\beta_{1,2}$ on a shot-by-shot basis from the intensities of four adjacent harmonics (SI Section VI). Such analysis reconstructs the joint statistics of $\sigma_{1,2}$, for each two-color delay.        
Figure 4a schematically describes the origin of the real and imaginary components of the sub-cycle perturbations to the action, highlighting the stochastic nature of the perturbation. The imaginary phase, $\beta_j$, corresponds to a modification of the tunneling, as the BSV modifies the ionization barrier near the peak of the coherent driving field, at half-cycle \(j\). The real term, $\alpha_j$, represents the additional phase accumulated by the electronic wavefunction between ionization and recombination, due to its interaction with the BSV field. 

\begin{figure}[!htb]
		\begin{center}
		
		\centering{\includegraphics*[width=0.95\columnwidth]{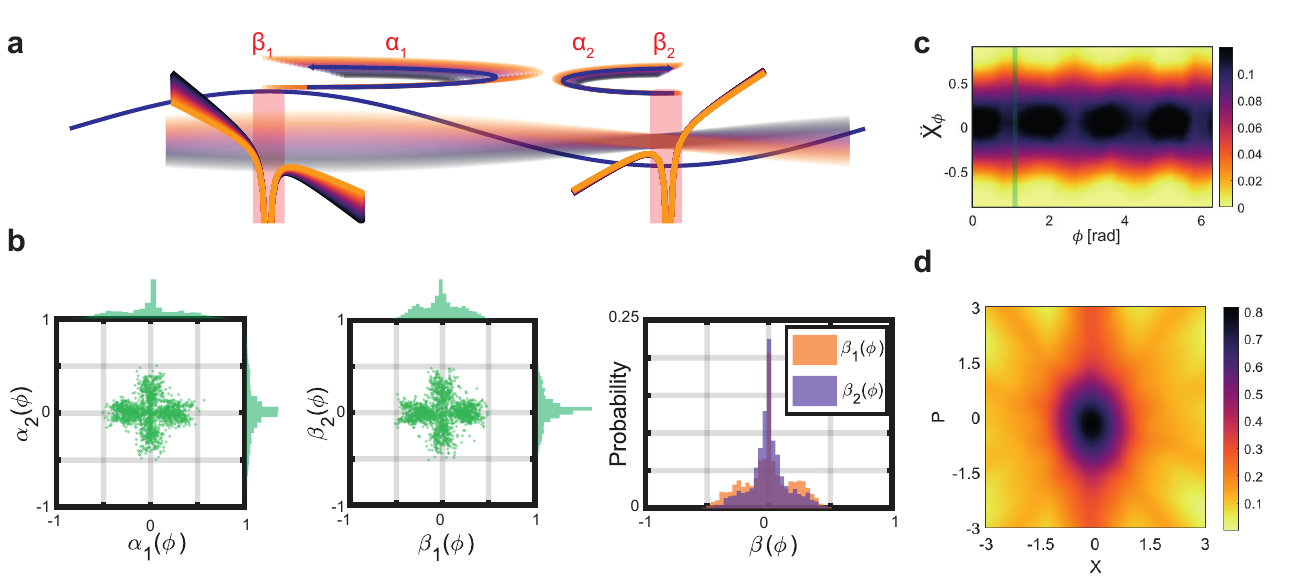}}
		\caption{ {\bf Quantum state tomography and sub-cycle electronic correlations.} {\bf a}, The coherent field (blue) induces pairs of trajectories, labeled 1 and 2, every half cycle of the fundamental field.  $\beta_1$ and $\beta_2$ correlate two instantaneous tunneling events: when one tunneling event exhibits excess noise (resulting in a broader $\beta_1$ distribution), the subsequent event, occurring half a cycle later, experiences reduced noise (leading to a narrower $\beta_2$ distribution). $\beta_1$ and $\alpha_1$ correlate a tunneling event with the corresponding electron trajectory. $\alpha_1$ and $\alpha_2$ correlate two successive trajectories. Finally, $\alpha_1$ and $\beta_2$ correlate a trajectory and the subsequent tunneling event. {\bf b}, Correlation between the sub-cycle events, extracted at the two-color delay which maximizes quadrature variance. {\bf c}, Quadrature amplitude $X_\phi$ of harmonic 14.5 extracted from the correlated events presented in {\bf b}, as a function of the two-color delay. The green line corresponds to the  value of delay, described in \textbf{b}. {\bf d}, Inverse Radon transform of $X_\phi$, representing the Wigner function of harmonic 14.5, up to a scaling factor. A vanishing displacement and a squeezed correlation are observed.}
			\label{finalFigure}
		\end{center}
	\end{figure}

Performing single-shot measurements of the harmonic spectrum allows us to reconstruct the statistical distribution of these complex phases as well as their correlations along successive half-cycles.
For example, the $\beta_1,\beta_2$ correlation in Fig. \ref{finalFigure}b reveals that when one tunneling event exhibits excess noise (resulting in a broad $\beta_1$ distribution), the subsequent event, occurring half a cycle later, experiences reduced noise (leading to a narrower $\beta_2$ distribution). This pattern corresponds to the temporal interval between excess and reduced noise in the input BSV. 

Finally, we present quantum state tomography of XUV harmonics. Traditionally, quantum state tomography is realized via homodyne tomography, in which an electromagnetic field $E\equiv X+iP$  is combined with a local oscillator (LO) of varying phase $\phi$, enabling measurement of the quadrature $X_\phi=\Re{Ee^{-i\phi}}$ across different LO phases $\phi$. From these quadrature distributions, one can reconstruct the field’s Wigner function via an inverse Radon transform \cite{Smithey1993}. In our experiment, a homodyne-like procedure arises internally through the interference of multiple photonic pathways. 
For example, the amplitude and phase statistics of half-integer ($2N+ \frac{1}{2}$) harmonics is resolved by virtue of the interference of a single-BSV-photon process—whose complex field is $\sigma\equiv\sigma_2+i\sigma_1$ —and a three-BSV-photon process proportional to $\sigma^3$. The quadrature amplitude $X_\phi$ of these half-integer harmonics is derived to be proportional to the real part of $\sigma$ evaluated at a two-color phase $\phi$ (SI section VI). This two-color phase $\phi$ plays a role analogous to the local oscillator phase in traditional homodyne detection. 
Mathematically, as $\phi$ is scanned, $\sigma$ “rotates” in the complex plane from $\sigma$ at $\phi=0$ to $\sigma e^{-i\phi}$. Consequently, the observable reconstructed by our interferometer effectively measures a series of rotated quadratures of $\sigma$, mirroring a sub-cycle homodyne detection scheme, resolving the quantum states of half-integer harmonics.
Figure 4c shows the quadrature distribution $X_\phi$ of harmonic 14.5, with a negligible mean value relative to its standard deviation ($\max_{\phi} \langle X_\phi \rangle/\Delta X_\phi=0.13$)
and oscillating variance (quadrature variance ratio $\min_{\phi}{\Delta X^2_\phi}/\max_{\phi}{\Delta X^2_\phi}=0.55$). 
This reconstruction indicates vanishing field displacement and squeezed-like fluctuations. Based on this measurement, we reconstruct the harmonic’s Wigner function up to a scaling factor (Fig. \ref{finalFigure}d) by an inverse Radon transformation (SI section VI), revealing a squeezed state centered at zero displacement. 

To the best of our knowledge, this is the first experimental demonstration of quantum-state tomography in the XUV spectral range. This reconstruction highlights the application of self-referenced internal interferometry which serves as a viable substitute for conventional homodyne tomography at attosecond time scales. We note that this scheme is complementary to quantum state tomography of attosecond electrons \cite{priebe2017attosecond,bourassin2020quantifying,koll2022experimental,laurell2025measuring}. 
Isolating the squeezed half-integer harmonic can provide a well-defined quantum source for further applications.

\section{\NoCaseChange{Conclusion}}
In this paper, we extend quantum light control and measurement into the attosecond XUV regime. By integrating infrared BSV with a strong coherent field, we imprint the squeezed fluctuations of the driving light onto both sub-cycle ionization dynamics and emitted harmonics. Scanning the two-color delay precisely controls the relative phase between photonic pathways, modulating both the mean intensity and photon statistics of the high harmonics. Our interferometric scheme directly reveals sub-cycle dynamics driven by the non-classical field. We report the first observation of tunneling driven by squeezed light, demonstrating that sub-cycle ionization fluctuations reflect the quantum correlations of the driving field. Our measurements show that tunneling statistics acquire a squeezed character, establishing a direct link between quantum optical noise and the tunneling mechanism. The interferometric scheme enables quantum state tomography — widely established in the visible and infrared regime — for the first time in the XUV regime. Our results bridge quantum optics and attosecond science, opening a path toward generating, characterizing, and controlling XUV pulses with non-classical properties. 
Non-classical attosecond physics holds great potential for quantum-enhanced metrology, interferometry, and beyond, laying the foundation for attosecond-scale quantum electrodynamics (QED) where electron and photon quantum states can be manipulated with unprecedented temporal precision.

\newpage{}

\bibliographystyle{naturemag}
\bibliography{references}

\section*{\NoCaseChange{Acknowledgments}}
This project has received funding from the Israel Science Foundation (ISF) under grant agreements Nos. 1315/24 and 2626/23. M.E.T. gratefully acknowledges the support of the Council for Higher Education scholarship for excellence in quantum science and technology. N.D. is the incumbent of the Robin Chemers Neustein Professorial Chair. N.D. acknowledges the Minerva Foundation, the Israeli Science Foundation and the European Research Council for financial support. M.E.T., M.B., I.N., I.K., M.K., and  O.C. thank the Helen Diller Quantum Center for partial financial support. We thank Maria Chekhova for discussions and Yossi Pilas for his technical support.


	%


\end{document}